\documentclass[
]{ceurart}

\usepackage{listings}
\usepackage{inputenc}
\usepackage{graphicx}
\usepackage{caption}
\usepackage{subcaption}
\graphicspath{ {./} }
\begin{document}

\copyrightyear{2022}
\copyrightclause{Copyright for this paper by its authors.
  Use permitted under Creative Commons License Attribution 4.0
  International (CC BY 4.0).}

\conference{MediaEval'22: Multimedia Evaluation Workshop,
  January 13--15, 2023, Bergen, Norway and Online}

\title{Identifying Misinformation Spreaders: A Graph-Based  Semi-Supervised Learning Approach}

% \tnotetext[1]{You can use this document as the template for preparing your publication. We recommend using the latest version of the ceurart style.}

\author[1]{Atta Ullah}[%
    % orcid=<orcid>,
    email=attaullah@cs.qau.edu.pk,
    % url=<website>,
]%\fnmark[1]

\author[1]{Rabeeh Ayaz Abbasi}[%
    email=rabbasi@qau.edu.pk,
]\cormark[1]%\fnmark[1]

\author[1]{Akmal Saeed Khattak}[%
    email=akhattak@qau.edu.pk,
]%\fnmark[1]

\author[2]{Anwar Said}[%
    email=anwar.said@vanderbilt.edu,
]%\fnmark[1]

% Affiliations 
\address[1]{Department of Computer Science, Quaid-i-Azam University Islamabad, Pakistan}

\address[2]{Institute for Software Integrated Systems, Department of Computer Science, Vanderbilt University, USA}

\cortext[1]{Corresponding author.}
% \fntext[1]{These authors contributed equally.}

\begin{abstract}
In this paper we proposed a Graph-Based conspiracy source detection method for the MediaEval task 2022 FakeNews: Corona Virus and Conspiracies Multimedia Analysis Task. The goal of this study was to apply SOTA graph neural network methods to the problem of misinformation spreading in online social networks. We explore three different Graph Neural Network models: GCN, GraphSAGE and DGCNN. Experimental results demonstrate that DGCNN outperforms in terms of accuracy.

% Your readers will read your abstract in order to decide whether or not to read the entire paper.
% Make it short, but informative: include you task name, a characterization of your approach, and a very brief statement of the main insights your achieved.
\end{abstract}

\maketitle

\section{Introduction}\label{sec:intro}
Social platforms such as Facebook and Twitter are becoming popular sources for day-to-day news consumption due to their convenience, low cost, and fast spread. In August 2021,the Pew Research Center in America reported that \verb|72%| of adults are more involved in social media, and \verb|48%| get much of their news from such platforms. On the contrary, misinformation spreading has posed challenges, which primarily influences the lives of people. There is no way to check the authenticity of the fake news. However, automated detection of fake news on such platforms is quite challenging. Furthermore, owing to the extraordinary volume of information, distinguishing between fake and real news is almost impossible. With the advent of automated technologies, both academia and industry have piqued an interest in delivering solutions\cite{das2021heuristic}. During COVID-19 outbreak, the American Journal of Tropical Medicine and Hygiene discovered that approximately 5800 patients had been hospitalized due to misinformation spread on social media. Furthermore, plenty of people died from drinking methanol or drugs that contain alcohol. They were misinformed that such products are helpful in the treatment of COVID-19 virus\cite{biradar2022combating}. 

This study is based on a dataset used in the MediaEval challenge 2022 FakeNews \cite{mediatask}. The challenge consists of three sub-tasks: content-based, graph-based, and a combination of both. This study is focused on a graph-based solution for the detection of misinformation spreaders.

\section{Related Work}\label{sec:work}
Misinformation or fake news spreading helps exaggerate information which makes it difficult to distinguish fake news from real news. There are three ways to identify fake news: through content, context, or a combination of the two. In content-based, the underlying challenge is the fluctuating nature of style, patterns, topics and platforms. Models trained on one dataset may not perform well when using a different dataset due to the differences in their contents, style, or language. To address such challenges, context-based solutions have been devoted to the detection of misinformation spreaders. Because it is obvious that the propagation of fake news and real news are different\cite{das2021heuristic}.

In \cite{kipf2016semi} the authors have presented a semi-supervised model GCN for node classification, and \cite{sharma2021identifying} extends it for rumors detection. GCN models are inherently transductive and fail to generalize unseen nodes. Therefore, \cite{hamilton2017inductive} have proposed an approach called GraphSAGE, which is a general inductive framework and effectively leverages node feature information for generation of every new node's embeddings. Instead of training each node's embeddings, they learn to produce embeddings by aggregating features from a node's local neighborhood. For more promising results, we consider changing node classification to graph classification. In \cite{xu2018powerful} authors have proposed a method called Graph Isomorphic Network (GIN), which classifies both graphs and nodes. Despite the strength of graph representation learning, GNN has a limited understanding of representational properties and limitations. The model is powerful as the Weisfeiler-Lehman graph isomorphism test\cite{xu2018powerful}. Similarly, in \cite{zhang2018end} proposed a method named, Deep Graph Convolutional Neural Network (DGCNN) to capture the graph level features which consist of four layers of either \cite{kipf2016semi}, \cite{hamilton2017inductive} or \cite{xu2018powerful}.

\section{Approach}\label{sec:approach}
We present the different methods we implemented in this study. As working on sub-task 2\cite{mediatask}, we have chosen the problem as both node and graph classification.

In node classification, the model aims to classify a node by analyzing its features and its neighbors' features. Similarly, in the graph classification setting, we construct subgraphs for each node by taking ego-network of each labeled node up to 3-hop neighbours.The implemented methods are GCN\cite{kipf2016semi}, GraphSAGE\cite{hamilton2017inductive} and DGCNN\cite{zhang2018end}.

\subsection{Graph Convolutional Neural Network}
\begin{figure}[htbp]
  \centering
  \includegraphics[width=\linewidth]{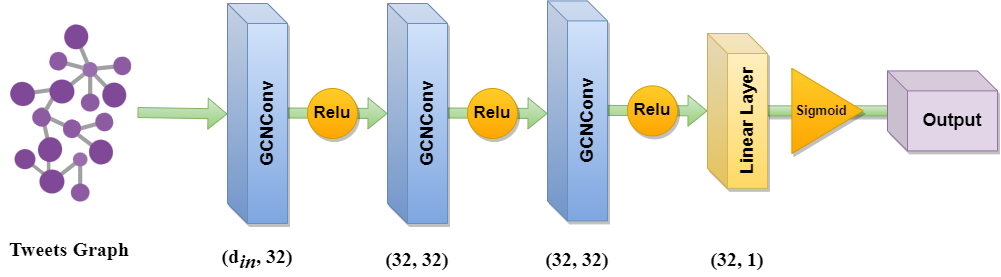}
  \caption{GCN architecture for user classification}
  \label{fig:GCN}
\end{figure}

Graph Convolutional Neural network (GCN) is a semi-supervised learning method for node classification. GCN is based on the message passing mechanism that learns from node's features and its neighboring node's features. We have used three GCN\cite{kipf2016semi} layers and the fourth is a linear layer. We use Binary Cross Entropy loss and Adam Optimizer during training. The architecture of the implemented model is illustrated in Figure \ref{fig:GCN}.

\subsection{GraphSAGE}
\begin{figure}[htbp]
\centering
\includegraphics[width=\linewidth]{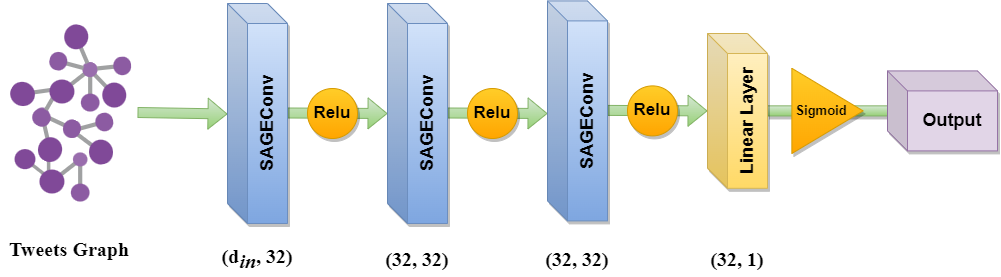}
\caption{GraphSAGE Architecture for User Classification}
\label{fig:GraphSage}
\end{figure}

We use three GraphSAGE layers along with the RELU activation function and a linear layer. Further, Binary Cross Entropy loss and Adam Optimizer during training has been utilized for measuring the performance of the proposed model. Figure \ref{fig:GraphSage} illustrates the architecture of the implemented GraphSAGE model. The rationale behind using GraphSAGE is: it is inductive and tries to create embeddings by using sampling and aggregation features from the node local neighborhood.

\subsubsection{Deep Graph Convolutional Neural Network}
DGCNN introduces a readout or pooling function which aggregates learned node embeddings to graph-level embedding\cite{biradar2022combating}.

We generate subgraphs for each node and transform the node classification problem to graph classification. The subgraphs are generated by taking the ego-network of up-to 3-hop neighbourhood of a node. This way we form the label mapping between node and corresponding ego-network of the node; the model classifies the label of a node's ego-network, which is taken as the label of the node.

\begin{figure}[htbp]
\centering
\includegraphics[width=\linewidth]{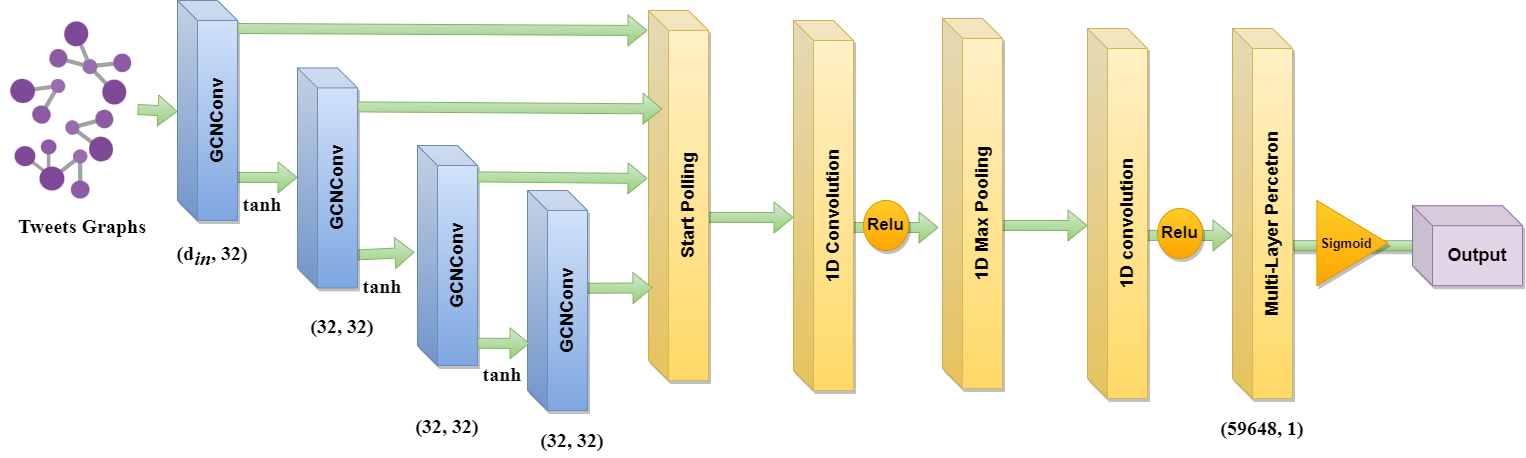}
\caption{DGCNN Architecture for User Classification}
\label{fig:DGCNN}
\end{figure}

The implemented model consists of four GCN layers, 1D-MaxPooling in between two 1D-Conv layers, and a fully connected layer as shown in Figure \ref{fig:DGCNN}.The network is trained using Binary Cross Entropy loss and the Adam optimizer with an initial learning rate of 0.$1e^-5$ and a dropout of 0.5.

\section{Results and Analysis}
In this section, we present our experimental results obtained through the implemented models. We use $80:20$ train and test split ratio and use accuracy, Matthews correlation coefficient (MCC) and Area Under the ROC (AUC) as evaluation metrics.

\begin{table}[htbp]
  \caption{Models' performance using different evaluation metrics}
  \label{tab:freq}
  \begin{tabular}{lccc}
    \toprule
    Model&Accuracy&MCC&ROC AUC\\
    \midrule
    GCN & \verb|60.54| &\verb|23.89|&\verb|57.54|\\
    GraphSage & \verb|65.79| &\verb|25.46|&\verb|67.02|\\
    DGCNN & \verb|69.85|  & \verb|36.24|&\verb|73.07|\\
  \bottomrule
\end{tabular}
\end{table}

We report models' performance in terms of three evaluation metrics in Table\ref{tab:freq} We conclude from the results that DGCNN performs quite better as compared to other two models.

\section{Discussion and Outlook}
As discussed the performance of implemented models in Section 4, these are the best results being obtained by using GCN, GraphSAGE, and DGCNN. The feature distribution of misinformation spreader and regular users is approximately equal through which the performance is considerably low. In order to improve results, combination of both Label Propagation (LPA) and GCN can help to classify misinformation spreaders and regular users effectively.

\def\bibfont{\small}
\bibliography{references}

\begin{thebibliography}{8}
\expandafter\ifx\csname natexlab\endcsname\relax\def\natexlab#1{#1}\fi
\providecommand{\url}[1]{\texttt{#1}}
\providecommand{\href}[2]{#2}
\providecommand{\path}[1]{#1}
\providecommand{\DOIprefix}{doi:}
\providecommand{\ArXivprefix}{arXiv:}
\providecommand{\URLprefix}{URL: }
\providecommand{\Pubmedprefix}{pmid:}
\providecommand{\doi}[1]{\href{http://dx.doi.org/#1}{\path{#1}}}
\providecommand{\Pubmed}[1]{\href{pmid:#1}{\path{#1}}}
\providecommand{\bibinfo}[2]{#2}
\ifx\xfnm\relax \def\xfnm[#1]{\unskip,\space#1}\fi
%Type = Inproceedings
\bibitem[{Das et~al.(2021)Das, Basak, and Dutta}]{das2021heuristic}
\bibinfo{author}{S.~D. Das}, \bibinfo{author}{A.~Basak},
  \bibinfo{author}{S.~Dutta},
\newblock \bibinfo{title}{A heuristic-driven ensemble framework for covid-19
  fake news detection},
\newblock in: \bibinfo{booktitle}{International Workshop onCombating Online
  Hostile Posts in Regional Languages during Emergency Situation},
  \bibinfo{organization}{Springer}, \bibinfo{year}{2021}, pp.
  \bibinfo{pages}{164--176}.
%Type = Article
\bibitem[{Biradar et~al.(2022)Biradar, Saumya, and
  Chauhan}]{biradar2022combating}
\bibinfo{author}{S.~Biradar}, \bibinfo{author}{S.~Saumya},
  \bibinfo{author}{A.~Chauhan},
\newblock \bibinfo{title}{Combating the infodemic: Covid-19 induced fake news
  recognition in social media networks},
\newblock \bibinfo{journal}{Complex \& Intelligent Systems}
  (\bibinfo{year}{2022}) \bibinfo{pages}{1--13}.
%Type = Misc
\bibitem[{MediaEval(2022)}]{mediatask}
\bibinfo{author}{MediaEval}, \bibinfo{title}{Fakenews: Corona virus and
  conspiracies multimedia analysis task}, \bibinfo{year}{2022}. \URLprefix
  \url{https://github.com/konstapo/2022-Fake-News-MediaEval-Task/}.
%Type = Article
\bibitem[{Kipf and Welling(2016)}]{kipf2016semi}
\bibinfo{author}{T.~N. Kipf}, \bibinfo{author}{M.~Welling},
\newblock \bibinfo{title}{Semi-supervised classification with graph
  convolutional networks},
\newblock \bibinfo{journal}{arXiv preprint arXiv:1609.02907}
  (\bibinfo{year}{2016}).
%Type = Inproceedings
\bibitem[{Sharma and Sharma(2021)}]{sharma2021identifying}
\bibinfo{author}{S.~Sharma}, \bibinfo{author}{R.~Sharma},
\newblock \bibinfo{title}{Identifying possible rumor spreaders on twitter: A
  weak supervised learning approach},
\newblock in: \bibinfo{booktitle}{2021 International Joint Conference on Neural
  Networks (IJCNN)}, \bibinfo{organization}{IEEE}, \bibinfo{year}{2021}, pp.
  \bibinfo{pages}{1--8}.
%Type = Article
\bibitem[{Hamilton et~al.(2017)Hamilton, Ying, and
  Leskovec}]{hamilton2017inductive}
\bibinfo{author}{W.~Hamilton}, \bibinfo{author}{Z.~Ying},
  \bibinfo{author}{J.~Leskovec},
\newblock \bibinfo{title}{Inductive representation learning on large graphs},
\newblock \bibinfo{journal}{Advances in neural information processing systems}
  \bibinfo{volume}{30} (\bibinfo{year}{2017}).
%Type = Article
\bibitem[{Xu et~al.(2018)Xu, Hu, Leskovec, and Jegelka}]{xu2018powerful}
\bibinfo{author}{K.~Xu}, \bibinfo{author}{W.~Hu},
  \bibinfo{author}{J.~Leskovec}, \bibinfo{author}{S.~Jegelka},
\newblock \bibinfo{title}{How powerful are graph neural networks?},
\newblock \bibinfo{journal}{arXiv preprint arXiv:1810.00826}
  (\bibinfo{year}{2018}).
%Type = Article
\bibitem[{Zhang et~al.(2018)Zhang, Cui, Neumann, and Chen}]{zhang2018end}
\bibinfo{author}{M.~Zhang}, \bibinfo{author}{Z.~Cui},
  \bibinfo{author}{M.~Neumann}, \bibinfo{author}{Y.~Chen},
\newblock \bibinfo{title}{An end-to-end deep learning architecture for graph
  classification},
\newblock \bibinfo{journal}{Proceedings of the AAAI Conference on Artificial
  Intelligence} \bibinfo{volume}{32} (\bibinfo{year}{2018}). \URLprefix
  \url{https://ojs.aaai.org/index.php/AAAI/article/view/11782}.
  \DOIprefix\doi{10.1609/aaai.v32i1.11782}.

\end{thebibliography}

\end{document}